\documentclass[a4paper,prb,twocolumn,showpacs,amsmath,amssymb]{revtex4}
\usepackage[english]{babel}
\usepackage[latin1]{inputenc}
\usepackage{amscd,verbatim}
\usepackage[dvips]{graphicx,color}
\usepackage[nooneline,tight,FIGTOPCAP,bf]{subfigure}
\usepackage{textcomp}
\usepackage{calc}
\usepackage{bbm}
\usepackage[OT1]{fontenc} 
\usepackage{natbib}
\usepackage[verbose,hypertexnames=false,bookmarksopenlevel=1,filecolor=blue,
linkcolor=blue,citecolor=blue,
bookmarksopen,
bookmarksnumbered,
colorlinks,
plainpages=false,
linktocpage
]{hyperref}

\newcommand{\bra}[1]{\left\langle #1\right|}
\newcommand{\ket}[1]{\left|#1\right\rangle}

\newcommand{\braOket}[3]{\left\langle #1\left|#2\right|#3\right\rangle}

\def\be{\begin{equation}}
\def\ee{\end{equation}}
\def\bea{\begin{eqnarray}}
\def\eea{\end{eqnarray}}
\def\bean{\begin{mathletters}\begin{eqnarray}}
\def\eean{\end{eqnarray}\end{mathletters}}

\def\Es2{E_{0,{\rm sat}}^2}

\newcounter{eqletter}
\def\mathletters{%
\setcounter{eqletter}{0}%
\addtocounter{equation}{1}
\edef\curreqno{\arabic{equation}}
\edef\@currentlabel{\theequation}
\def\theequation{%
\addtocounter{eqletter}{1}\thechapter.\curreqno\alph{eqletter}%
}%
}

\DeclareMathAlphabet{\mathpzc}{OT1}{pzc}{m}{it}

\DeclareSymbolFont{mathbold}{OML}{cmm}{b}{it}
\DeclareMathSymbol{\balpha}{\mathord}{mathbold}{11}
\DeclareMathSymbol{\bbeta}{\mathord}{mathbold}{12}
\DeclareMathSymbol{\bgamma}{\mathord}{mathbold}{13}
\DeclareMathSymbol{\bdelta}{\mathord}{mathbold}{14}
\DeclareMathSymbol{\bepsilon}{\mathord}{mathbold}{15}
\DeclareMathSymbol{\bvarepsilon}{\mathord}{mathbold}{34}
\DeclareMathSymbol{\bzeta}{\mathord}{mathbold}{16}
\DeclareMathSymbol{\bEta}{\mathord}{mathbold}{17}
\DeclareMathSymbol{\btheta}{\mathord}{mathbold}{18}
\DeclareMathSymbol{\bvartheta}{\mathord}{mathbold}{35}
\DeclareMathSymbol{\biota}{\mathord}{mathbold}{19}
\DeclareMathSymbol{\bkappa}{\mathord}{mathbold}{20}
\DeclareMathSymbol{\blambda}{\mathord}{mathbold}{21}
\DeclareMathSymbol{\bmu}{\mathord}{mathbold}{22}
\DeclareMathSymbol{\bnu}{\mathord}{mathbold}{23}
\DeclareMathSymbol{\bxi}{\mathord}{mathbold}{24}
\DeclareMathSymbol{\bpi}{\mathord}{mathbold}{25}
\DeclareMathSymbol{\bvarpi}{\mathord}{mathbold}{36}
\DeclareMathSymbol{\brho}{\mathord}{mathbold}{26}
\DeclareMathSymbol{\bvarrho}{\mathord}{mathbold}{37}
\DeclareMathSymbol{\bsigma}{\mathord}{mathbold}{27}
\DeclareMathSymbol{\bvarsigma}{\mathord}{mathbold}{38}
\DeclareMathSymbol{\btau}{\mathord}{mathbold}{28}
\DeclareMathSymbol{\bupsilon}{\mathord}{mathbold}{29}
\DeclareMathSymbol{\bphi}{\mathord}{mathbold}{30}
\DeclareMathSymbol{\bvarphi}{\mathord}{mathbold}{39}
\DeclareMathSymbol{\bchi}{\mathord}{mathbold}{31}
\DeclareMathSymbol{\bpsi}{\mathord}{mathbold}{32}
\DeclareMathSymbol{\bomega}{\mathord}{mathbold}{33}
\DeclareMathSymbol{\biGamma}{\mathord}{mathbold}{0}
\DeclareMathSymbol{\biDelta}{\mathord}{mathbold}{1}
\DeclareMathSymbol{\biTheta}{\mathord}{mathbold}{2}
\DeclareMathSymbol{\biLambda}{\mathord}{mathbold}{3}
\DeclareMathSymbol{\biXi}{\mathord}{mathbold}{4}
\DeclareMathSymbol{\biPi}{\mathord}{mathbold}{5}
\DeclareMathSymbol{\biSigma}{\mathord}{mathbold}{6}
\DeclareMathSymbol{\biUpsilon}{\mathord}{mathbold}{7}
\DeclareMathSymbol{\biPhi}{\mathord}{mathbold}{8}
\DeclareMathSymbol{\biPsi}{\mathord}{mathbold}{9}
\DeclareMathSymbol{\biOmega}{\mathord}{mathbold}{10}

\newcommand{\kt}[1]{\text{#1}}

\newcommand{\I}{\mathit{i}}

\begin{document}
\title{Spin Hall Conductivity on the Anisotropic Triangular Lattice}

\author{P. Wenk}
\email[]{p.wenk@jacobs-university.de}
\homepage[]{www.physnet.uni-hamburg.de/hp/pwenk/}
\affiliation{School of Engineering and Science, Jacobs University Bremen, Bremen 28759, Germany}
\author{S. Kettemann}
\email[]{s.kettemann@jacobs-university.de}
\homepage[]{www.jacobs-university.de/ses/skettemann}
\affiliation{School of Engineering and Science, Jacobs University Bremen, Bremen 28759, Germany, and Division of Advanced Materials Science Pohang University of Science and Technology (POSTECH) San31, Hyoja-dong, Nam-gu, Pohang 790-784, South Korea}
\author{G. Bouzerar}
\email[]{georges.bouzerar@grenoble.cnrs.fr}
\homepage[]{neel.cnrs.fr}
\affiliation{Institut N\'{e}el, CNRS, Grenoble Cedex 9 38042, France and School of Engineering and Science, Jacobs University Bremen, Bremen 28759, Germany}

\begin{abstract}
We present a detailed study of the spin Hall conductivity on a two-dimensional triangular lattice in the presence of Rashba spin-orbit coupling. In particular,  we focus part of our attention on the effect of the anisotropy of the nearest neighbor hopping amplitude. It is found that the presence of anisotropy has drastic effects on the spin Hall conductivity, especially in the hole doped regime where a significant increase or/and reversed sign of the spin Hall conductivity has been obtained.  We also provide a systematic analysis of the numerical results in terms of Berry phases. The changes of signs observed at particular density of carriers appear to be a consequence of both Fermi surface topology and change of sign of electron velocity. In addition, in contrast to the two-dimensional square lattice, it is shown that the tight binding spin-orbit Hamiltonian should be derived carefully from the continuous model on the triangular lattice.
\end{abstract}
\maketitle
\section{Introduction}
The incorporation of electron spin degree of freedom and its associated magnetic moment into electronic devices has initiated a rapidly growing field of spin electronics, called spintronics. One of the first idea of spintronic device is attributed to Datta and Das in 1990\cite{Datta1990a}. The authors have proposed a spin field-effect transistor with the innovation of manipulating a pure spin current. Such a manipulation is possible due to the relativistic effect of the coupling between spin and orbital degrees of freedom which can be manipulated e.g. by means of a gate voltage. This spin-orbit coupling (SOC) can emerge in various ways in semiconductors. One source is the lack of inversion symmetry of the confining potential, which defines the 2D electron gas, generated by the heterostructure of the semiconductor. This Bychkov-Rashba\cite{rashba_1,rashba_2} type SOC has the advantage that it can be directly tuned by an applied gate voltage. Another way to introduce SOC in the sample is by choosing a 
noncentrosymmetric material where the lack of inversion center creates a SOC called Dresselhaus\cite{Dresselhaus1955} SOC.
Developing further the correspondence between (charge) electronics and spintronics with the aim of creating pure spin currents, D'yakonov and Perel \cite{Dyakonov1971.13} proposed already in 1971 the spin Hall effect (SHE). This mechanism requires spin dependent impurity scattering such as skew scattering or side jump mechanism\cite{Schliemann2006a}. In today's terminology, it is known as extrinsic SHE. It was experimentally confirmed in 2004/2005 by angle-resolved optical detection of spin polarization at the edges of a two-dimensional layer.\cite{PhysRevLett.94.047204,Kato2004}

In this paper, we will focus on the SHE which arises even in the absence of impurities, the so called intrinsic SHE, which is due to SOC. The latter is assumed in this paper to be of Bychkov-Rashba type. It lifts, without breaking time reversal symmetry, the spin degeneracy of the eigenstates. The theory of the SHE has been developed in the last ten years, as reviewed in Refs.\,\onlinecite{Schliemann2006a,Engel2007}, but only recently experimental evidence has been reported for electrical manipulation and detection of intrinsic SHE in ballistic HgTe/HgCdTe quantum wells.\cite{Brune2010}\\
Starting at low electron fillings in the conduction band where both Rashba bands are occupied, one finds the ``universal`` spin Hall conductivity (SHC) $\sigma_\kt{SH}=e/8\pi$, which is independent of the SOC strength for sufficiently small couplings.\cite{Sinova2004, WenkThesis}
Going beyond the long wavelength approximation, the underlying lattice geometry becomes important.
On a square lattice using a single-orbital model (spin-split) the SHC $\sigma_\kt{SH}$ shows electron-hole symmetry: $\sigma_\kt{SH}$ is an odd function of the Fermi energy $E_F$ which thus vanishes at half-filling.\cite{Sinova2004,WenkThesis} However both, shape and sign of SHC are lattice structure dependent. 
This can, for instance, nicely be illustrated with the appearance of several plateaus as seen in the SHC calculations on honeycomb and kagome lattice by Liu \textit{et al.}, Ref.\,\onlinecite{Liu20092091,PhysRevB.79.035323}.
Both of these lattices contain several atoms per unit cell respectively two and three. The kagome lattice can be seen as a triangular lattice with 3 atoms per unit cell. The question which arises is whether similar differences as seen between square and kagome SHC are expected already in the case of the simple triangular lattice. In contrast to the square lattice, the triangular lattice brings along the property of geometrical frustration which is at the origin of exotic properties. In addition,  electron-hole symmetry is broken on the triangular lattice. Among exciting properties, the triangular based lattices are known to favor in some cases ferromagnetic groundstates when the electron-electron interaction is switched on.\cite{PhysRevB.54.4056,JPSJ.66.2123,PhysRevB.81.014421}
This is also known as flat band ferromagnetism.\cite{mielke1991,mielke1992,PhysRevLett.69.1608,MielkeTasaki1993} In these systems, the intrinsic frustrating nature of the lattice is a crucial ingredient in the stabililization of ferromagnetism. The importance of geometrical frustration has also been shown 
to play an important role in superfluid-Mott insulator quantum phase transitions in ultracold quantum gases.\cite{1367-2630-12-6-065025} It has been suggested that the supersolid phase reported results from the competition between Mott localization and frustration.\cite{PhysRevLett.95.127207} Among other interesting features found on triangular lattices one can also mention superconductivity in CoO$_2$ layered compounds\cite{Takada2003} or various anisotropies in organic bisethylenedithio conductors\cite{doi:10.1021/cr030646k}.\\

Back to the SHE, it is interesting to mention that there are already experimental realizations of surfaces with triangular geometry.
 Indeed, giant Rashba SOC has been recently reported on surfaces of Au$(111)$, Bi$(111)$ and Bi/Ag$(111)$.\cite{PhysRevLett.77.3419} 
In this manuscript we present a detailed analysis of the SHC on the anisotropic triangular lattice as a function of the carrier concentration. In addition, the obtained numerical results will be discussed in the context of Berry phases\cite{Berry08031984}. One of the aims of including anisotropy in the hopping amplitudes is to allow us to tune in a continuous way the position of the Van Hove singularity and in the same time change the topology of the Fermi surface. Experimentally, the anisotropy in the hopping integrals could be achieved by introducing strains or growing samples on surfaces which have small 
differences in the lattice parameters.

\section{Tight Binding Hamiltonian on Triangular Lattice with SOC}
\begin{figure}[t]
\begin{center}
\includegraphics[width=\columnwidth*2/3]{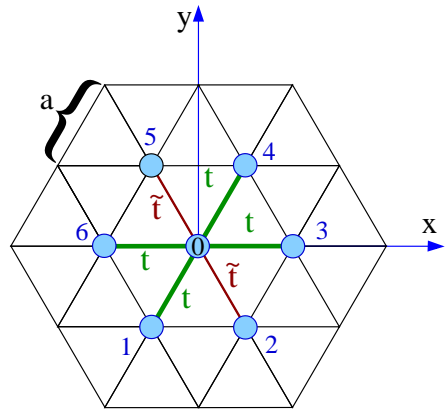}
   \caption{(Color online) Two dimensional triangular lattice in the x-y plane. $t$ and $\tilde t$ are the hopping amplitudes, $a$ is the lattice constant. The directions are numbered.}\label{plot:latticeDefinition}
    \end{center}
\end{figure}

We consider a 2D electron gas on the triangular lattice as sketched in Fig.\,\ref{plot:latticeDefinition}.

Including the Rashba SOC with coupling strength $\alpha_2$ the Hamiltonian reads
\begin{align}
 H=&{}\frac{\hat{k}^2}{2m^*}+\alpha_2(\hat{k_y}\sigma_x-\hat{k_x}\sigma_y)\\
=&{}\frac{1}{2m^*}({\bf \hat{k}}+e{\bf A_{SO}})^2-m^*\alpha_2^2\label{Aso},
\end{align}
where,
\begin{align}
 {\bf A_{SO}}=&{}\frac{2m^*}{e}\hat a{\bf S},\\
\hat a=&{}\left(\begin{matrix}
           0 & -\alpha_2\\
	  \alpha_2 & 0
          \end{matrix}\right),
\end{align}
and $\sigma_i$, $i=x,y$ are the Pauli matrices, ${\bf S}$ is the spin-$1/2$ operator and  $m^*$ the  electron effective mass.

As it will become clear in the following, it is convenient to introduce the lattice unitary vectors,
\begin{align}
 {\bf e_x}=&{}\left(\begin{matrix}
           1\\
	   0
          \end{matrix}\right),\qquad
 {\bf e_4}=&{}\left(\begin{matrix}
           \frac{1}{2}\\
	   \frac{\sqrt{3}}{2}
          \end{matrix}\right),\qquad
 {\bf e_5}=&{}\left(\begin{matrix}
           -\frac{1}{2}\\
	   \frac{\sqrt{3}}{2}
          \end{matrix}\right).
\end{align}
The Hamiltonian can be recast in the following form,
\begin{align}
H=&{}\frac{2}{3}\frac{1}{2m^*}(({\bf e_x\cdot \hat{\tilde k}})^2+({\bf e_4\cdot \hat{\tilde k}})^2+({\bf e_5\cdot \hat{\tilde k}})^2)-m^*\alpha_2^2 \\
=&{}\frac{2}{3}\frac{1}{2m^*}(\hat{k}_x^2+\hat{k}_4^2+\hat{k}_5^2)-\frac{\alpha_2}{3}(\hat{k}_4-\hat{k}_5+2\hat{k}_x)\sigma_y\nonumber\\
{}&+\frac{\alpha_2}{\sqrt{3}}(\hat{k}_4+\hat{k}_5)\sigma_x\label{H_TBM1},
\end{align}
where the canonical moment is given by
\begin{align}
 {\bf\hat{\tilde k}}=&{}{\hat{\bf k}}+\alpha_2 m^* \left(\begin{matrix}
           -\sigma_y\\
	   \sigma_x
          \end{matrix}\right).
\end{align}
 The discretization of the Hamiltonian leads to the Tight Binding Hamiltonian (TBH) whose nearest neighbor hoppings are listed in Tab.\,\ref{tab:TBH}.  This Hamiltonian is identical to that one would obtain directly using the well known tight-binding expression of the SO part and that reads,  
\begin{align}\label{HR_general}
H^{(R)}={}&\frac{2}{3}\frac{\alpha_2}{2a}\sum_{\langle ij \rangle \sigma\sigma^\prime}\left(\bsigma\times \frac{{\hat{\bf e}}_{ij}}{\I}\right)_{z,\sigma,\sigma^\prime} c^\dagger_{i\sigma}c_{j\sigma^\prime},
\end{align}
where ${\hat{\bf e}}_{ij}$ is the unitary vector pointing from site $i$ and $j$.
Note the presence of the 2/3  coefficient in both Eq.\,(\ref{H_TBM1}) and in Eq.\,(\ref{HR_general}). We could absorb this coefficient in the hopping definition but we  have chosen to keep them in order to facilitate the direct comparison to existing calculations.  
The comparison between our TBH and that derived in Ref.\,\onlinecite{PhysRevB.76.075322} reveals differences in the hopping amplitudes. The difference in the derived TBH results from the discretization procedure. By using the relation between  $k_x$, $k_y$ and $k_4$, $k_5$ one can generate an infinite number of rigorously equivalent Hamiltonians in the continuous picture with different coefficients for $\sigma_x$ and $\sigma_y$. However, the discretization will lead to an infinite number of inequivalent TBHs. Thus this crucial step has to be done carefully. Note that this problem will not occur on the square lattice because the connecting vectors $e_x$ and $e_y$ are orthogonal to each other and the discretization can be done unambiguously. Then, how to proceed in order to derive the correct TBH?
There are two procedures. The most straightforward is to use directly Eq.\,(\ref{HR_general}) as often done in the literature. The second procedure consist in re-expressing the continuum Hamiltonian in term of the canonical momentum as done in Eq.\,(\ref{H_TBM1}) and then only discretize it by properly defining the real space derivative.
\begin{table}\label{tab:TBH}
\begin{center}
\large
\begin{tabular}{ccc}
\hline
\hline
$(i,j)$ & $H^{(0)}_{(i,j)}$ & $H^{(R)}_{(i,j)}$ \\
\hline
$(0,1)$ & $-\frac{2}{3}t_0 $ & $\I\frac{t^R}{\sqrt{3}}\sigma_x-\I\frac{t^R}{3}\sigma_y$ \\
$(0,2)$ & $-\frac{2}{3}t_0 $ & $\I\frac{t^R}{\sqrt{3}}\sigma_x+\I\frac{t^R}{3}\sigma_y$ \\
$(0,3)$ & $-\frac{2}{3}t_0 $ & $\I\frac{2 t^R}{3}\sigma_y$ \\
$(0,4)$ & $-\frac{2}{3}t_0 $ & $-\I\frac{t^R}{\sqrt{3}}\sigma_x+\I\frac{t^R}{3}\sigma_y$ \\
$(0,5)$ & $-\frac{2}{3}t_0 $ & $-\I\frac{t^R}{\sqrt{3}}\sigma_x-\I\frac{t^R}{3}\sigma_y$ \\
$(0,6)$ & $-\frac{2}{3}t_0 $ & $-\I\frac{2 t^R}{3}\sigma_y$\\
\hline
\end{tabular}
\caption{Matrix elements in the tight-binding model for the kinetic part $H^{(0)}$ and the Rashba part $H^{(R)}$, where $t^R=\alpha_2/(2a)$, $t_0=1/(2m^*a^2)$ and $a$ denotes the lattice constant.}
\end{center}
\end{table}
\subsection*{Spectrum and Density of States}
\begin{figure}[t]
\begin{center}
\includegraphics[width=\columnwidth*2/3]{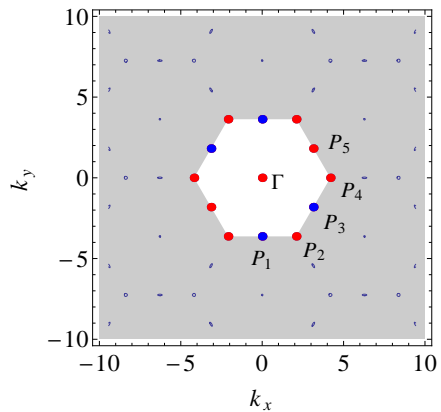}
   \caption{(Color online) Degeneracy points in the energy spectrum of $H$ in $k$-space (for the 1st Brillouin zone shown): $M\equiv P_1=\{0,-2\pi/\sqrt{3}\}^T$, $P_2=\{2\pi/3,-2\pi/\sqrt{3}\}^T$, $P_3=\{\pi,-\pi/\sqrt{3}\}^T$, $K\equiv P_4=\{4\pi/3,0\}^T$,  $P_5=\{\pi,\pi/\sqrt{3}\}^T$. The blue dots indicate the saddle points in case of vanishing SOC.}\label{plot:kSpaceDegeneracies}
    \end{center}
\end{figure}

Let us now proceed with the diagonalization of the anisotropic TBH. We define the hopping amplitude in the $x$ and $0-4$ direction as $t=(2/3)t_0$ and that in the $0-5$ direction as $\tilde{t}=(2/3)\tilde{t}_0$. 
From now on, the lattice constant $a$ will be set to one, $a\equiv 1$.\\
The spectrum of the TBH is given by
\begin{align}
 E_{\pm}({\bf k})={}&-\frac{2}{3}\left(2t_0\left(\cos(k_x)+\cos\left(k_4\right)\right)\right.\nonumber\\
{}&\left. +2\tilde t_0\cos\left(k_5\right)\pm\sqrt{2}t^R F_1({\bf k})\right),\label{spectrum}
\end{align}
where $F_1({\bf k})$ is defined as
\begin{align}
 F_1({\bf k})\equiv{}& \left(3+\cos(k_x)-\cos(2 k_x)\right.\nonumber\\
{}&-(1+2\cos(k_x))\cos(\sqrt{3} k_y)\nonumber\\
{}&\left.+8\cos\left(\frac{k_x}{2}\right)\cos\left(\frac{\sqrt{3} k_y}{2}\right)\sin\left(\frac{k_x}{2}\right)^2\right)^{\frac{1}{2}}.
\end{align}
The corresponding eigenvectors are
\begin{align}
\ket{\lambda^{\pm}({\bf k})}={}& \frac{1}{\sqrt{2}}
\left(
\begin{matrix}
 \pm e^{\I \phi({\bf k})}\\
  1
\end{matrix}
\right),\label{EV_general}
\intertext{with $\phi({\bf k})$ given by}
e^{\I \phi({\bf k})}={}&\frac{\I}{\sqrt{2}}F_1({\bf k})\left(\cos\left(\frac{\sqrt{3} k_y}{2}\right) \sin\left(\frac{k_x}{2}\right)\right.\nonumber\\
{}&\left.+\sin(k_x)+\I\sqrt{3}\cos\left(\frac{k_x}{2}\right) \sin\left(\frac{\sqrt{3}k_y}{2}\right)\right)^{-1}.
\end{align}
Before proceeding with the calculation of the SHC, let us first discuss some features of the energy spectrum which, as will be seen, will appear to be important and helpful to understand the characteristics of this dynamical quantity. As it is well known, the SHC can be directly related to the $k$-space curvature,\cite{PhysRevB.71.085315} this will be presented in detail in what follows, Sec.\,\ref{sec:Geometr_interpret}. In this context, the singular values of the Berry connection at degeneracy points of the energy spectrum are of crucial importance. To prepare the insight into the topological nature of this problem, we first calculate the degeneracies by solving directly $F_1(k)=0$. The result is plotted in Fig.\,\ref{plot:kSpaceDegeneracies}.

\begin{figure}[t]
 \begin{center}
   \subfigure[\qquad\qquad\qquad\qquad\qquad\qquad]{\scalebox{0.74}{\includegraphics[width=\columnwidth*2/3]{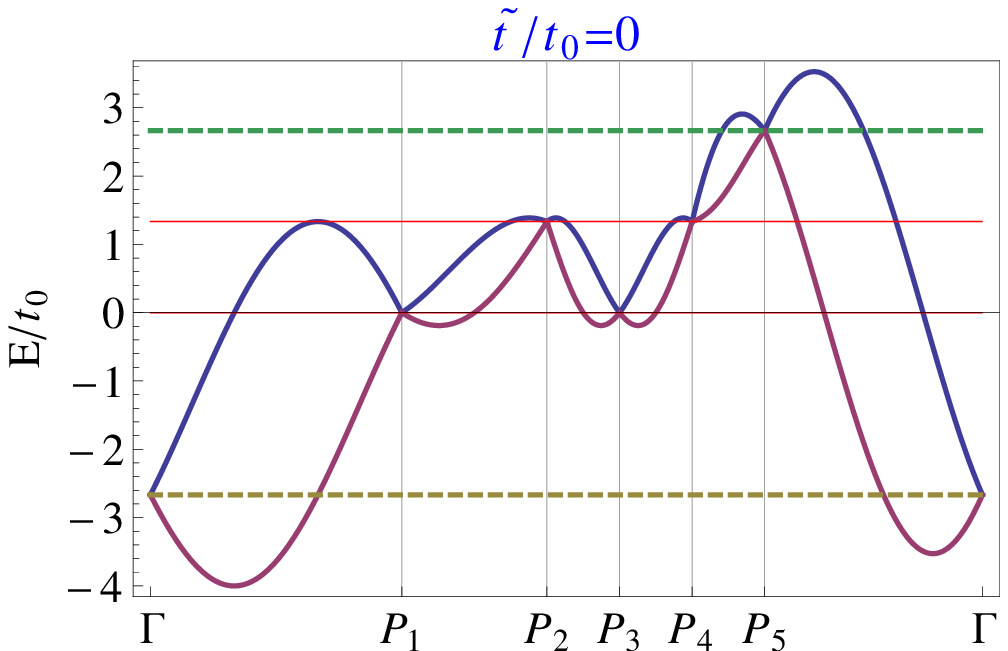}}}
   \subfigure[\qquad\qquad\qquad\qquad\qquad\qquad]{\scalebox{0.74}{\includegraphics[width=\columnwidth*2/3]{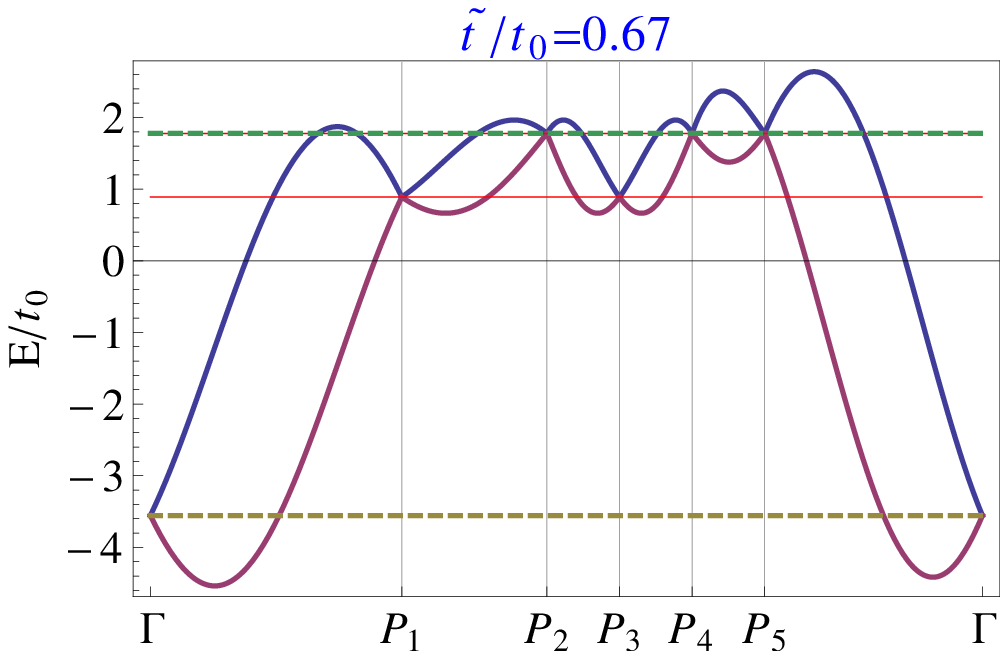}}}\\
   \subfigure[\qquad\qquad\qquad\qquad\qquad\qquad]{\scalebox{0.74}{\includegraphics[width=\columnwidth*2/3]{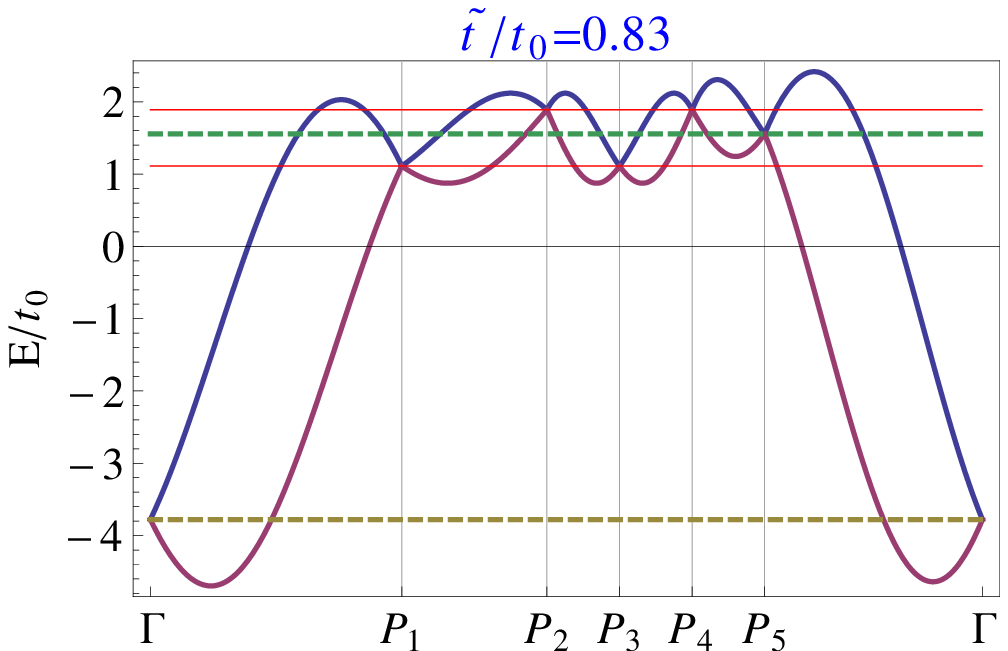}}}
   \subfigure[\qquad\qquad\qquad\qquad\qquad\qquad]{\scalebox{0.74}{\includegraphics[width=\columnwidth*2/3]{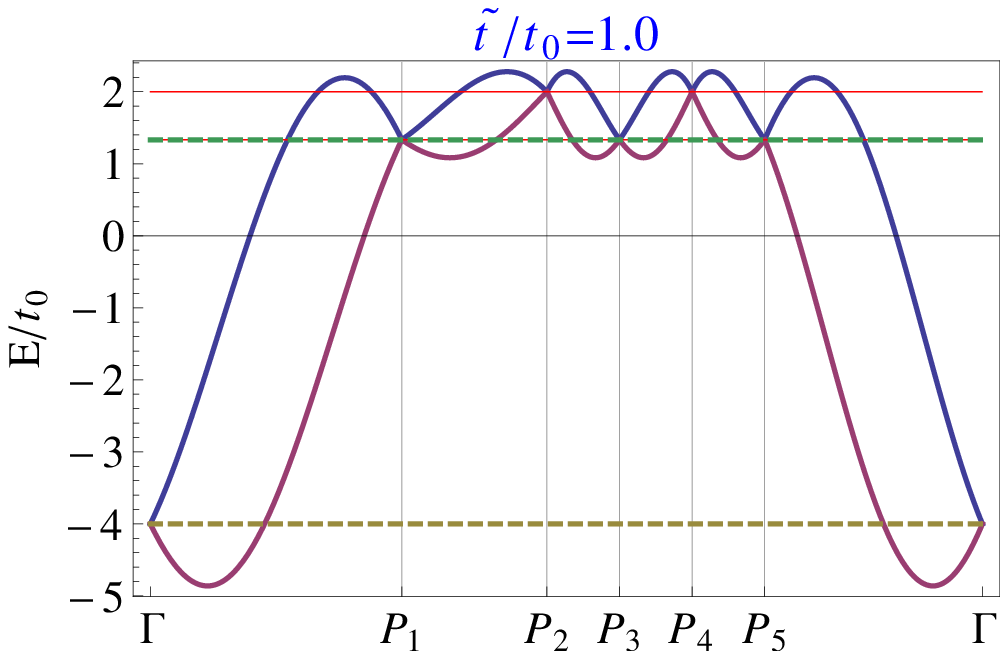}}}
   \caption{(Color online) The energy dispersion, Eq.\,\ref{spectrum}, is plotted for both modes $E_\pm$ (red curve: $E_-$). The $k$-points $P_i$ are according to Fig.\,\ref{plot:kSpaceDegeneracies}. The two red horizontal lines indicate $E_{S1}$ and $E_{S2}$. The green-dashed one indicates $E_{d2}$ and the orange-dashed $E_{d1}$. We have used a large SOC coupling, $t^R=1t_0$, for sake of clarity. (See text)}\label{plot:spectrum_degeneracies}
  \end{center}
\end{figure}

\begin{figure}[t]
 \begin{center}
\includegraphics[width=\columnwidth]{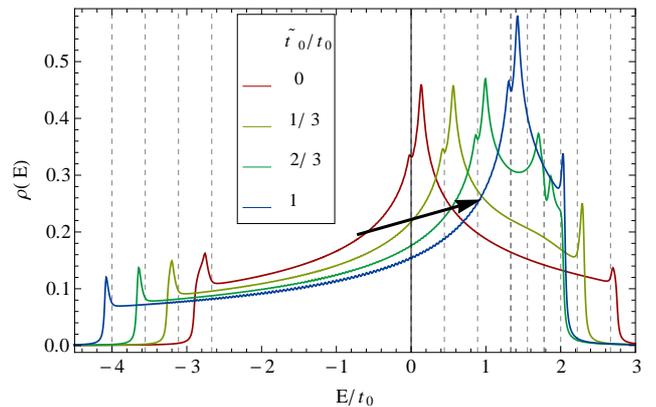}
   \caption{(Color online) DOS $\rho(E)$ for a system of size $L=300$ (cutoff in the Lorenzian was chosen to be $\eta=0.03$) and $t^R=0.3 t_0$, for different values of $\tilde t_0/t_0=0,0.2,\ldots, 1$. Dashed lines indicate the degeneracy points energy, see Fig.\,\ref{plot:spectrum_degeneracies}.}\label{plot:DOS_t-tild_var}
  \end{center}
\end{figure}

In the absence of SOC ($t^R$=0) and in the isotropic case (${t_0}=\tilde{t}_0$), the density of states (DOS) exhibits a Van Hove singularity at Fermi energy $E=4/3t_0$. It corresponds to a straight line which goes through $P_3$ and $P_5$. For finite values of SOC strength, in the isotropic case, this peak is now split as clearly seen in Fig.\,\ref{plot:DOS_t-tild_var}. 
In addition, one has two additional peaks at the boundary of the energy spectrum.\cite{Moca2008} The energy dispersion is plotted as a function of momentum in Fig.\,\ref{plot:spectrum_degeneracies}(d). As seen, the degeneracy points correspond to the points $P_i$ ($i=1,2,\ldots, 5$) and $\Gamma$.
One finds $E(P_1)=E(P_3)=E(P_5)$ and $E(P_2)=E(P_4)$. In Fig.\,\ref{plot:DOS_t-tild_var} these particular energies are indicated by dashed lines.
When we switch on the anisotropy the singularities in the DOS change significantly as expected. In Fig.\,\ref{plot:spectrum_degeneracies} one now sees that one degeneracy is lifted $E(P_1)=E(P_3) \neq E(P_5)$, and we still have $E(P_2)=E(P_4)$. Thus in the presence of anisotropy we have four relevant energies that we denote,
\begin{equation}\label{PiEnergies}
\begin{aligned}
 E(\Gamma)\equiv E_{d1}={}&-(4/3)(2+\tilde t_0/t_0),\\
 E(P_5)\equiv E_{d2}={}&(4/3)(2-\tilde t_0/t_0),\\
 E(P_1)=E(P_3)\equiv E_{S1}={}&(4/3)\tilde t_0/t_0,\\
 E(P_2)=E(P_4)\equiv E_{S2}={}&(4/3)+(2/3)\tilde t_0/t_0,
\end{aligned}
\end{equation}
with the energy expressed in values of $t_0$.
Note that, two different sets of labels ($d_1$,$d_2$) and ($S_1$, $S_2$) are introduced. The meaning of this separation will become clear in what follows.
\section{Spin Hall Conductivity}
\subsection*{Numerical Results of SHC}
Having analyzed the energy spectrum, we now proceed further with the calculation of dynamical properties, more precisely the spin Hall conductivity $\sigma_\kt{SH}$.
The SHC  is calculated within linear response theory using Kubo formula,\cite{JPSJ.12.570,PhysRevB.31.3372,Nomura2005,Moca2008}
\begin{align}
\sigma_{\kt{SH}}(E_F)={}& 2\frac{e}{V}\sum_{E_m<E_F<E_n}\frac{\Im(\bra{\psi_m}J_{x}^z\ket{\psi_n}\bra{\psi_n}v_{y}\ket{\psi_m})}{(E_n-E_m)^2+\eta^2},\label{SHC}
\end{align}
where $V$ is the total number of lattice sites, $E_i$ the eigenenergies, $\ket{\psi_i}$ the corresponding eigenstates, $E_F$ the Fermi energy and $\eta$ a positive infinitesimal. The spin current operator is
\begin{align}
 {\bf J}^z={}&\frac{\hbar}{4}\{\sigma_z,{\bf v}\},\label{j_classic}
\end{align}
and the velocity is defined with the usual expression ${\bf v}=\I[{\bf H,r}]$.
To calculate the SHC we first compute the matrix element density function $j(x,y)$ defined by
\begin{align}
{}&j(x,y)=\sum_{m,n}M_{mn}\delta(x-E_m)\delta(y-E_n),
\end{align}
where the matrix element $M_{mn}$ is
\begin{align}
 M_{mn}=\Im[\bra{\psi_m}J_{x}^z\ket{\psi_n}\bra{\psi_n}v_{y}\ket{\psi_m}].
\end{align}
Note that the function $j(x,y)$ is a particularly useful and convenient quantity for calculations carried out by Kernel Polynomial Method\cite{RevModPhys.78.275} in the presence of disorder.\cite{WenkThesis,KPM1}

\begin{figure}[t]
 \begin{center}
   \subfigure[]{\scalebox{0.85}{\includegraphics[width=\columnwidth]{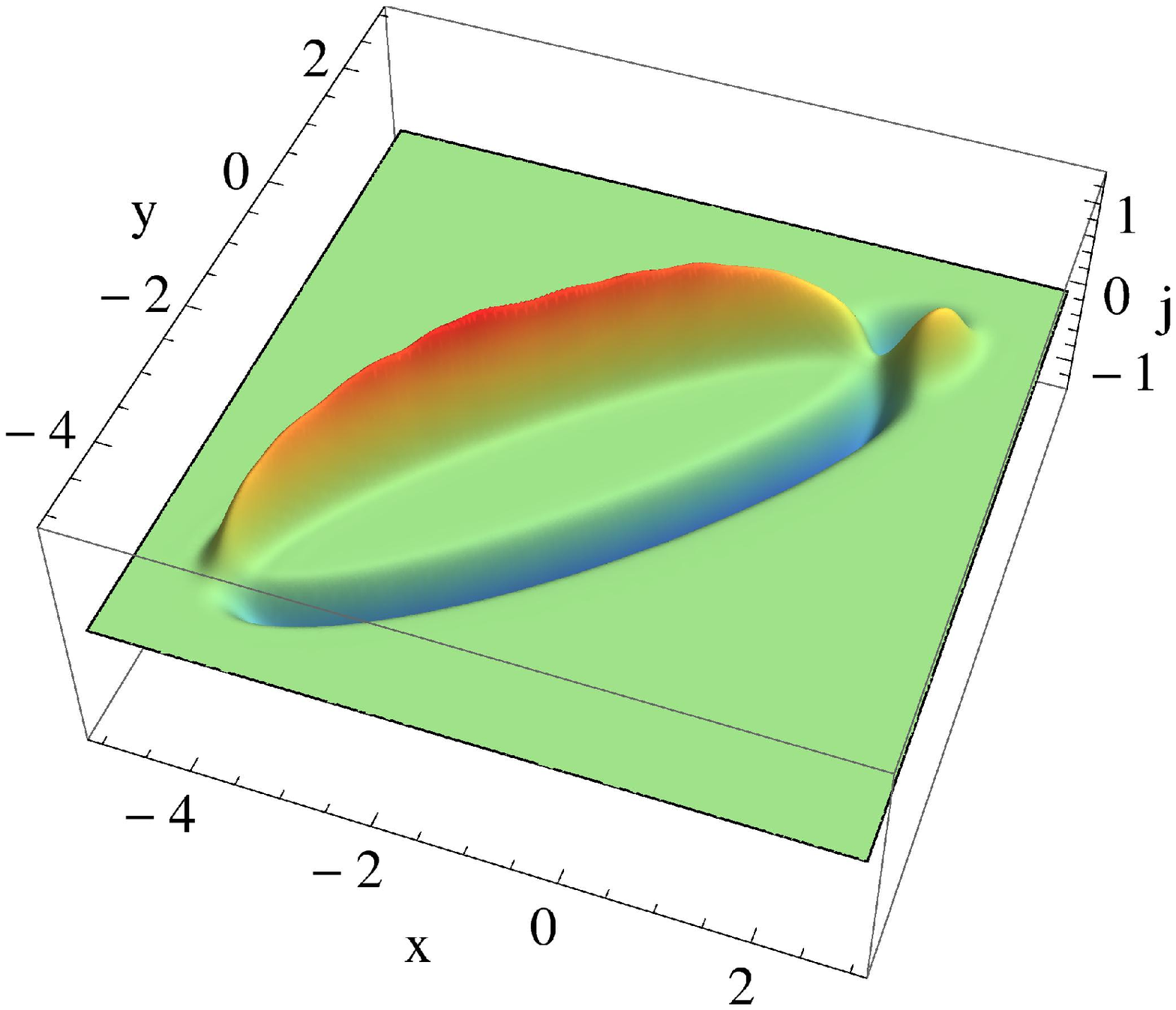}}}
   \subfigure[]{\scalebox{0.85}{\includegraphics[width=\columnwidth*3/4]{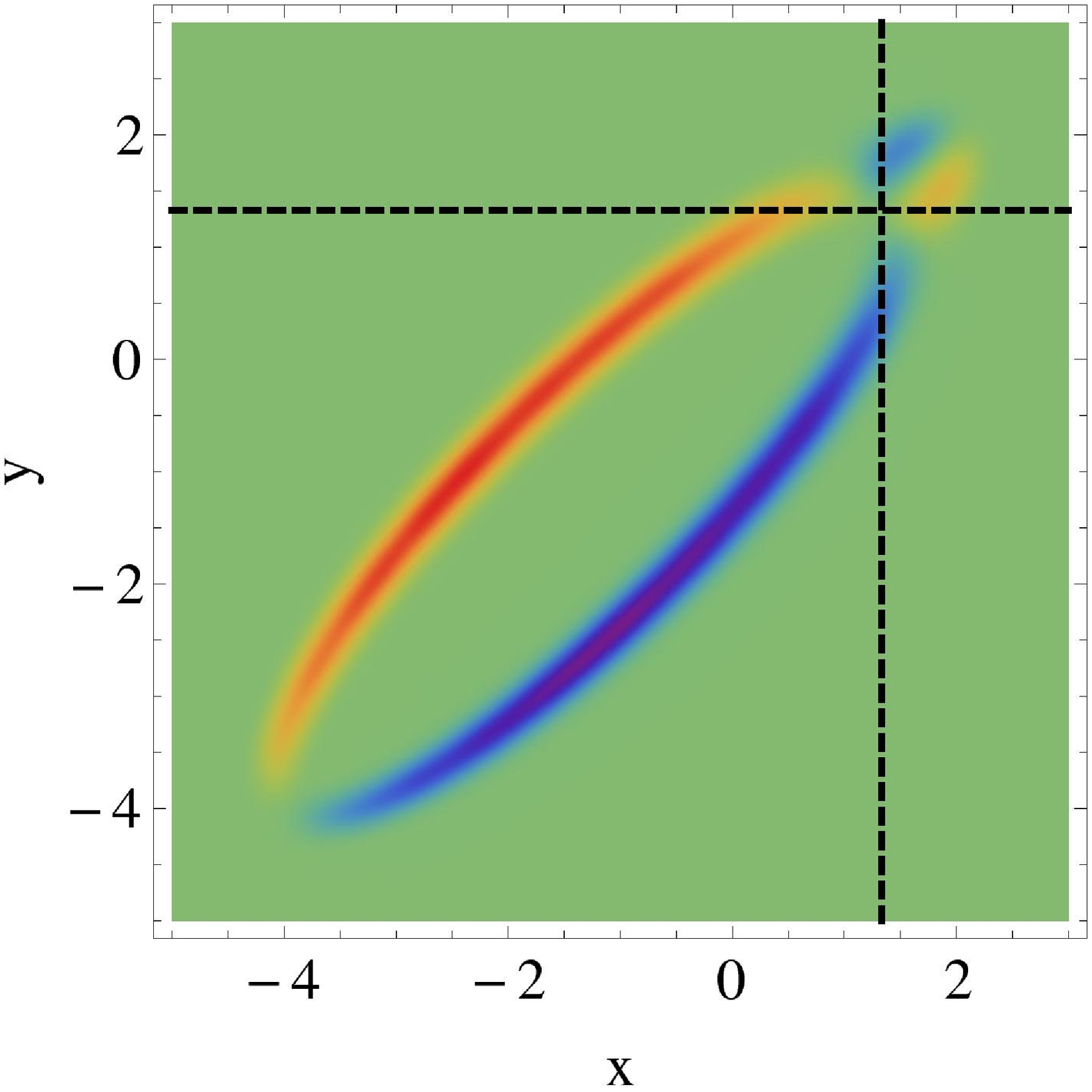}}}
   \caption{(Color online) Matrix element density function $j(x,y)$ for the isotropic case ($t_0=\tilde t_0$) [(b) is a 2D plot of (a)]. The system size is of $70\times 70$ sites. The SOC strength is $t^R=0.3t_0$. The dashed lines in (b) mark the position of $E_{S1}$.}\label{plot:jxy}
  \end{center}
\end{figure}

We can thus rewrite $\sigma_{\kt{SH}}(E_F)$ in terms of $j(x,y)$, leading to the following expression \cite{WenkThesis}:
\begin{align}
 {}&\sigma_{\kt{SH}}(E_F)=\nonumber\\ 
{}&\frac{e}{V} \int_{E_\kt{min}}^{E_\kt{max}}\int_{E_\kt{min}}^{E_\kt{max}} dx\,dy\,\frac{f_{E_F}(x)-f_{E_F}(y)}{(y-x)^2+\eta^2}j(x,y),
\end{align}
where $f_{E_F}$ is the Fermi-Dirac distribution.

\begin{figure}[t]
\begin{center}
   \subfigure[\!\!\!\!\!\!\!\!\!\!\!\!\!\!\!]{\scalebox{0.9}{\includegraphics[width=\columnwidth]{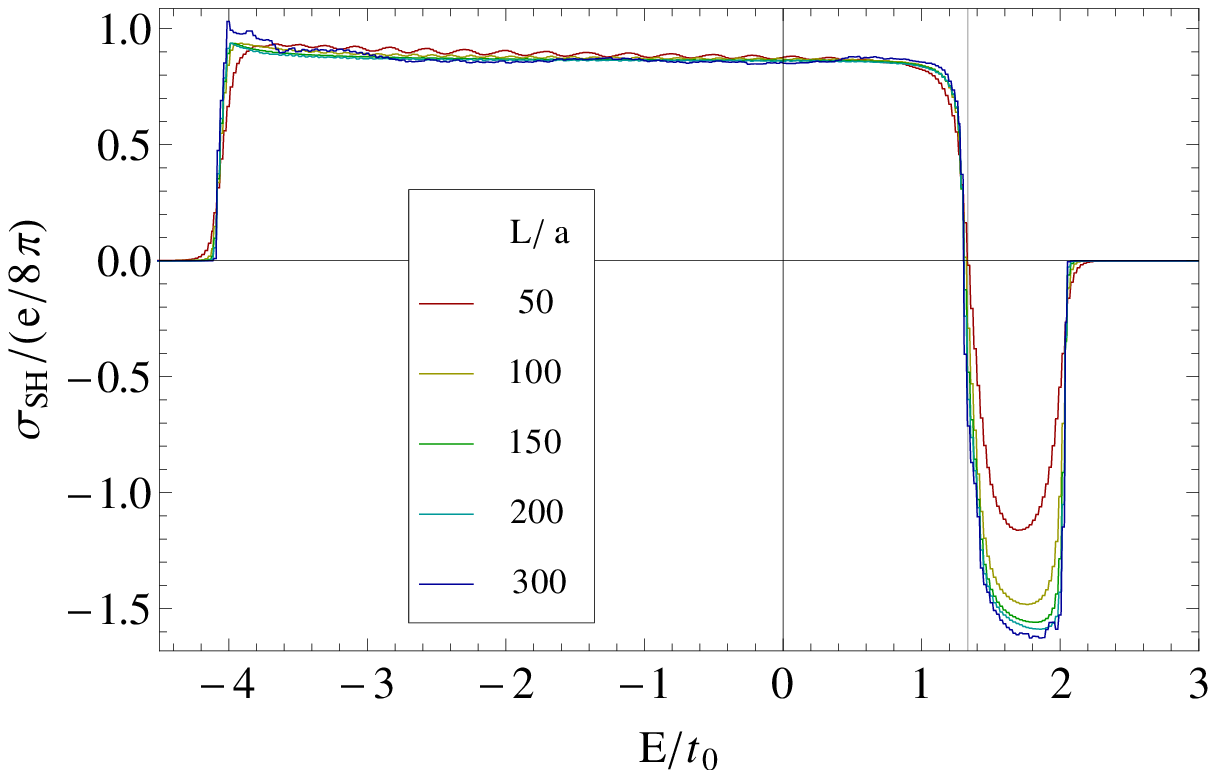}}}
   \subfigure[\!\!\!\!\!\!\!\!\!\!\!\!\!\!\!]{\scalebox{0.85}{\includegraphics[width=\columnwidth]{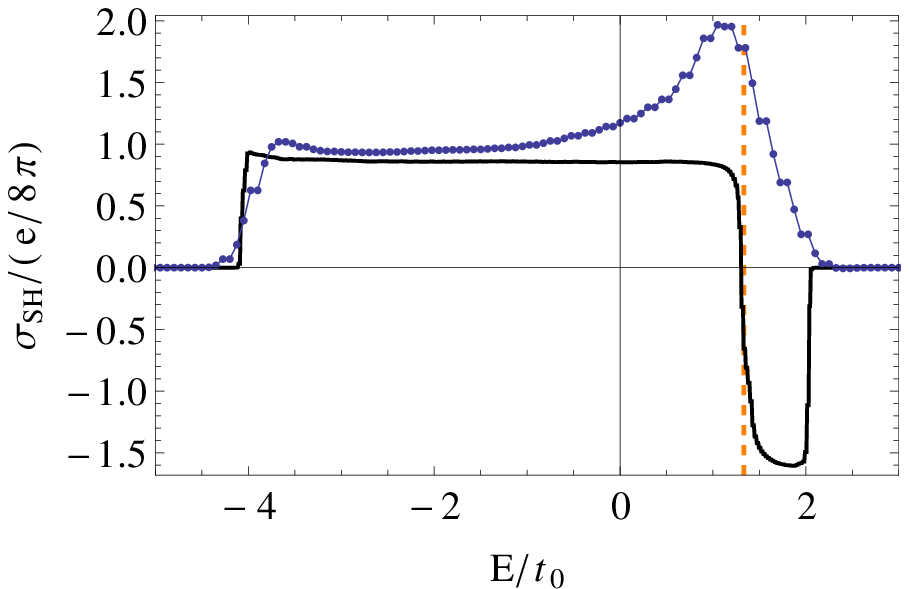}}}
   \caption{(Color online) (a) SHC for different system sizes $L=50,\ldots,300$ for the isotropic case. The SOC strength is $t^R=0.3t_0$. (b) Comparison between the $\sigma_{\kt{SH}}$ calculated within our model (continuous line) and that of Ref.\,\onlinecite{PhysRevB.76.075322} (filled circles). The system contains $70\times 70$ sites. The vertical dashed line corresponds to $E_{S1}$.}\label{plot:SHC_TriangularPRB76_075322}
    \end{center}
\end{figure}

In the present case the only non-zero-matrix elements are 
\begin{align}
M_{\pm}({\bf k})={}& -M_{\mp}({\bf k})\label{M_symmetry}\\
={}&\Im\{\bra{\lambda^+({\bf k})}J_{x}^z\ket{\lambda^-({\bf k})}\bra{\lambda^-({\bf k})}v_{y}\ket{\lambda^+({\bf k})}\}, 
\end{align}
where
\begin{align}
{}& M_{\pm}({\bf k})= 4
\sqrt{2} t^R \sin(k_x)\left(1+2\cos\left(\frac{k_x}{2}\right)\cos\left(\frac{\sqrt{3}k_y}{2}\right)\right)\nonumber\\
{}& \times\left(2t_0\sin(k_x)+t_0\sin\left(k_4\right)-\tilde t_0\sin\left(k_5\right)\right)/(3F_1({\bf k})).
\end{align}
The function $j(x,y)$ is plotted in Fig.\,\ref{plot:jxy} for the isotropic case. Notice the symmetry due to Eq.\,\ref{M_symmetry}.
One can clearly see a sign change for $j$ values at $(x,y>x)$ resp. $(x,y<x)$. This gives a clear indication for a sign-change in $\sigma_\kt{SH}$ at a particular Fermi level. In fact, this energy coincides with $E_{S1}$.
The resulting SHC is plotted in Fig.\,\ref{plot:SHC_TriangularPRB76_075322}. At low electron density we observe a sharp increase to $\sigma_\kt{SH}=e/(8\pi)$ as expected since the lattice structure is irrelevant. As we increase the Fermi energy further, $\sigma_\kt{SH}$ remains almost flat up to $E/t_0 \approx 1$ (extended plateau).
Then at $E=E_{S1}$ the sign of SHC changes in agreement with the discussion on $j(x,y)$. More remarkable is the change of amplitude of the SHC in the region of negative conductivity. Indeed, it increases by 60$\%$ with respect to the case of low electron density.

\begin{figure}[htbp]
\begin{center}
   \includegraphics[width=\columnwidth]{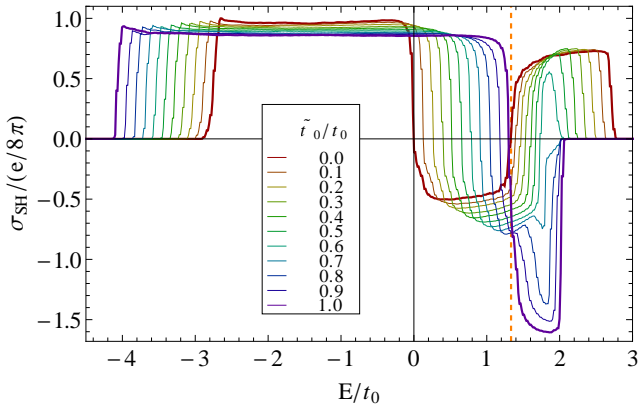}
   \caption{(Color online) SHC for $t^R=0.3t_0$ and $\tilde t_0/t_0={0, 0.1,\ldots, 1}$. The local maximum at low energies is shifting with $E_{d1}$. The sign-change at low Fermi energy is found at $E_{S1}$, the other one (in the case of $\tilde t_0/t_0<2/3$) is found at $E_{S2}$.  The vertical dashed line corresponds to $E_{S1}(\tilde t_0=t_0)=E_{S2}(\tilde t_0=0)$.}\label{plot:SHC_all_ttild0}
    \end{center}
\end{figure}

Interestingly, the calculations indicate that the finite size effects are almost negligible up to $E_{S1}$, but not in the hole doped regime. As mentioned before, one has to be careful in the derivation of TBH. We now compare the calculated SHC with that obtained using the TBH of Ref.\,\onlinecite{PhysRevB.76.075322}. The result is plotted in Fig.\,\ref{plot:SHC_TriangularPRB76_075322} (b). As seen, the results are relatively comparable up to $E_F=-2t_0$ and drastically deviate in the vicinity of $E_{S1}$.
The conductivity calculated within the TBH of Ref.\,\onlinecite{PhysRevB.76.075322} exhibits a clear maximum in the vicinity of $E_{S1}$, but no change of sign.
We now analyze the effect of anisotropy in the hopping amplitude on the SHC. The results are depicted in Fig.\,\ref{plot:SHC_all_ttild0}.
As we discussed it previously a change of sign occurs at $E_{S1}$.  As we reduce $\tilde t_0$ we observe a shift of the crossing energy towards $E_F=0$. One also notices drastic changes in the shape of $\sigma_{\kt{SH}}$ beyond $E_{S1}$. In particular, we observe two minima and a significant reduction of the amplitude of SHC. This is especially pronounced for the case $\tilde t_0=0.7 t_0$. As the anisotropy is further decreased, the second minimum disappears and a second change of sign of the SHC is found. This occurs at the value of $\tilde t_0=(2/3) t_0$. Thus, below this value one has two changes of sign, one at $E_{S1}$ and the second at $E_{S2}$. One natural question which arises is why beyond this value SHC changes its sign a single time? This will be answered and understood when discussing the SHC in the framework of Berry phases.
Remark that the second change of sign could not be anticipated from the DOS. Indeed, in contrast to the peak seen at $E_{S1}$ no peak, no singularity or pronounced feature is visible at $E_{S2}$, see for instance the case $\tilde t_0=(1/3) t_0$ in Fig.\,\ref{plot:DOS_t-tild_var}. The limiting case 
  $\tilde t_0=0$ requires also some additional attention. In the absence of SOC, the system is topologically equivalent to the square lattice. As seen in Fig.\,\ref{plot:SHC_all_ttild0} the results for $\sigma_{\kt{SH}}$ are similar to that of the square lattice for $E_F<0$ (see e.g. Ref.\,\onlinecite{Sheng2005,PhysRevB.75.035325,WenkThesis}). However, for $E_F>0$ (below $E_F/t_0<1$), the amplitude is reduced to half of that of the square lattice.
The spin flip term in the $(2-5)$ direction at finite SOC is at the origin of these  crucial differences in the hole doped regime.
It is surprising that the drastic changes take place only for $E_F>0$, we will shed light on it in what follows.
\begin{figure}[t]
\begin{center}
\includegraphics[width=\columnwidth*3/4]{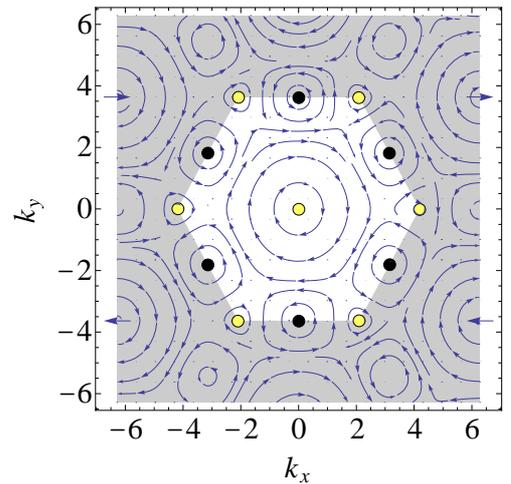}
   \caption{Berry connection ${\bf \mathcal{A}^\pm(k)}$ shown as a stream plot in ${\bf k}$-space. Divergences of the Berry connection $\mathcal{A}$ are found at the degeneracy points of the energy spectrum indicated by small circles. Clockwise/counterclockwise winding of $\mathcal{A}$ around this degeneracies is indicated by yellow/black color of this circles.}\label{plot:Berry_connection}
    \end{center}
\end{figure}
\subsection*{Geometric Interpretation of SHC}\label{sec:Geometr_interpret}
In order to get a better understanding of the previously discussed features, we propose to analyze the results in the context of the Berry phase.
As could be guessed from comparing the spectra shown in Fig.\,\ref{plot:spectrum_degeneracies} and the SHC plotted in Fig.\,\ref{plot:SHC_all_ttild0}, the topology of the spectrum plays an important role in understanding the transport properties of electrons and their spins. We remind that the transport of a Bloch state $\ket{\lambda(\bzeta)}$ in the parameter space $\bzeta$ which has a non-vanishing curvature leads to picking up a geometrical phase. In the case of adiabatic transport, valid in the linear response regime, it is called Berry phase.\cite{Berry08031984} In momentum space this geometrical phase is at the origin of the quantum,\cite{PhysRevLett.49.405,JPSJ.62.659,Mahito1985343} the anomalous\cite{PhysRevLett.83.3737} and spin\cite{Murakami05092003,PhysRevB.69.235206} Hall effects.  The latter was also studied in the context of Fermi-surface topologies on honeycomb and kagome lattice by Liu {\it et al.} in Ref.\,\onlinecite{Liu20092091,PhysRevB.79.035323}.\\ 
In our case the Berry phases are explicitly given by the integral
\begin{align}
 \gamma^\pm=\oint_{C^{\pm}}{\bf \mathcal{A}^\pm(k)}\cdot d{\bf k},
\end{align}
along a loop $C^\pm=\{k_x,k_y\}$ in momentum space with the condition $E^\pm(k_x,k_y)=E_F$, the Berry connection ${\bf \mathcal{A}^\pm(k)}$ is
\begin{align}\label{Berry_connection}
 {\bf \mathcal{A}^\pm(k)}={}&\braOket{\lambda^\pm({\bf k})}{\frac{\partial}{\I\partial{\bf k}}}{\lambda^\pm({\bf k})}.
\end{align}

\begin{figure}[t]
\begin{center}
   \subfigure[\!\!\!\!\!\!\!\!\!\!\!\!\!\!\!]{\scalebox{0.85}{\includegraphics[width=\columnwidth*4/7]{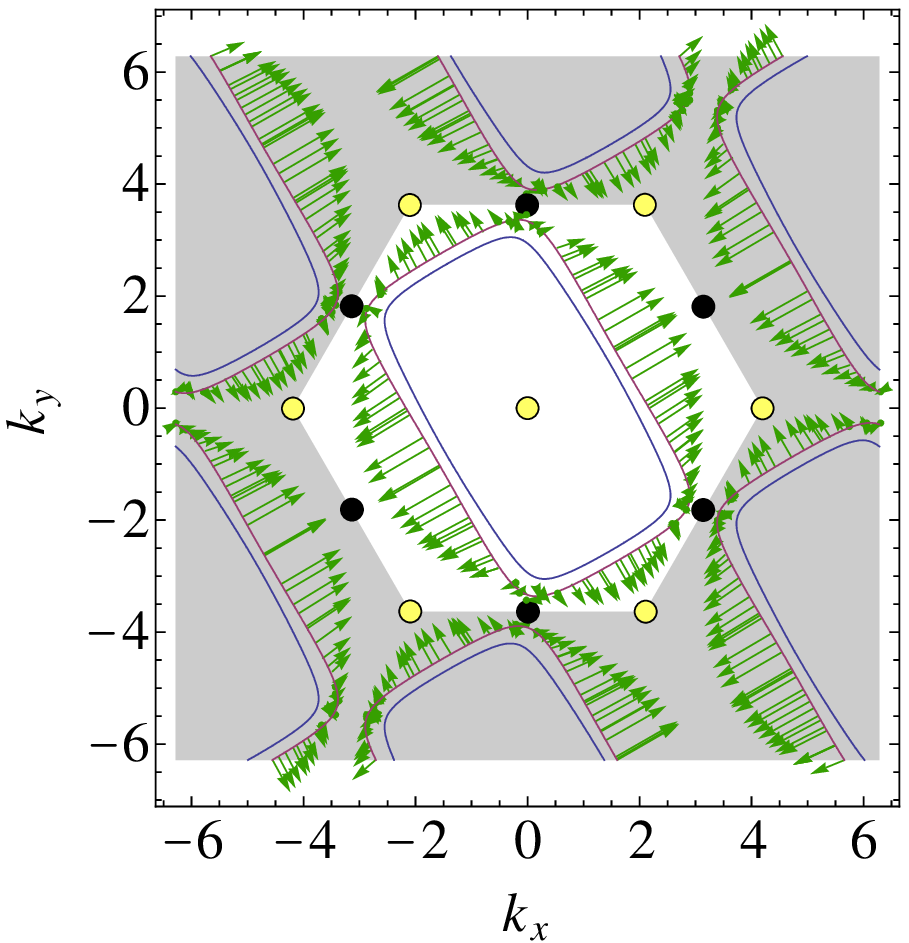}}}
   \subfigure[\!\!\!\!\!\!\!\!\!\!\!\!\!\!\!]{\scalebox{0.85}{\includegraphics[width=\columnwidth*4/7]{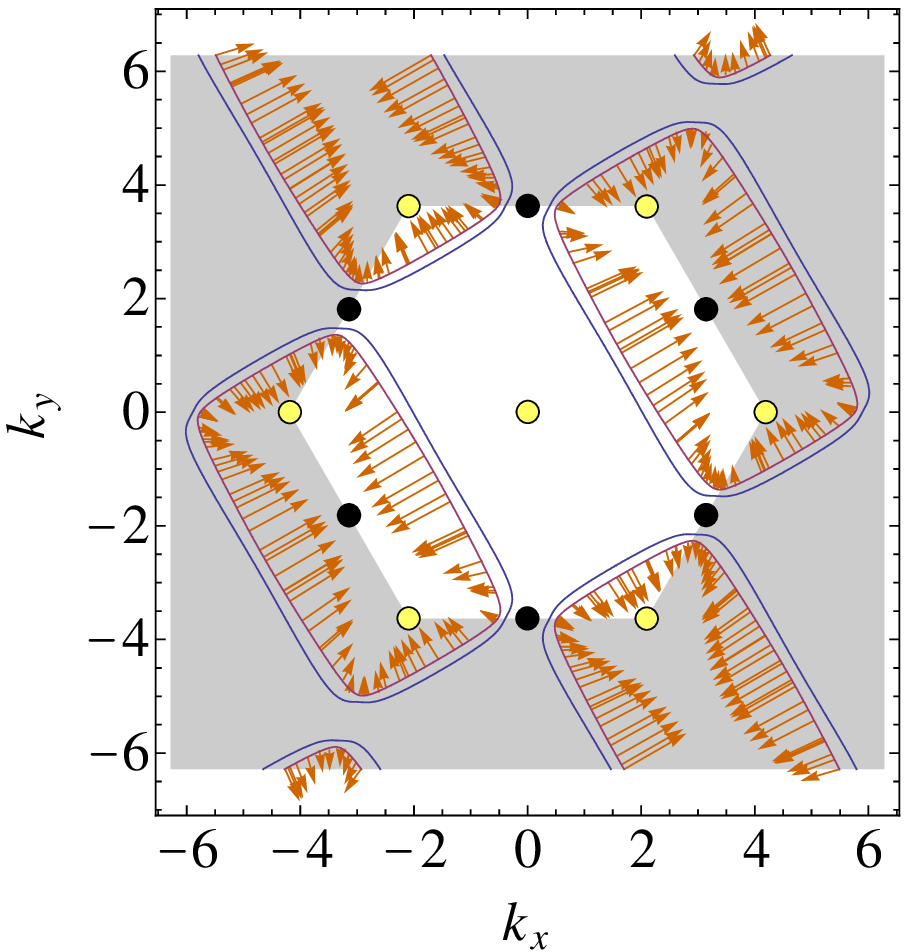}}}
   \subfigure[\!\!\!\!\!\!\!\!\!\!\!\!\!\!\!]{\scalebox{0.85}{\includegraphics[width=\columnwidth*4/7]{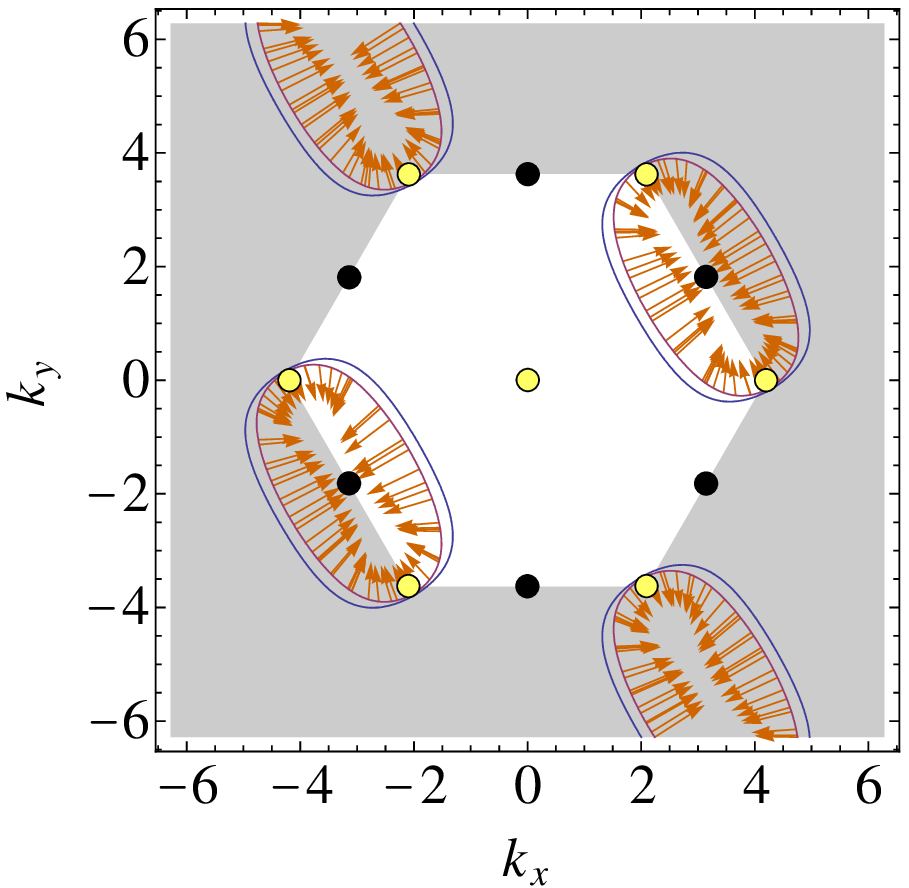}}}
   \subfigure[\!\!\!\!\!\!\!\!\!\!\!\!\!\!\!]{\scalebox{0.85}{\includegraphics[width=\columnwidth*4/7]{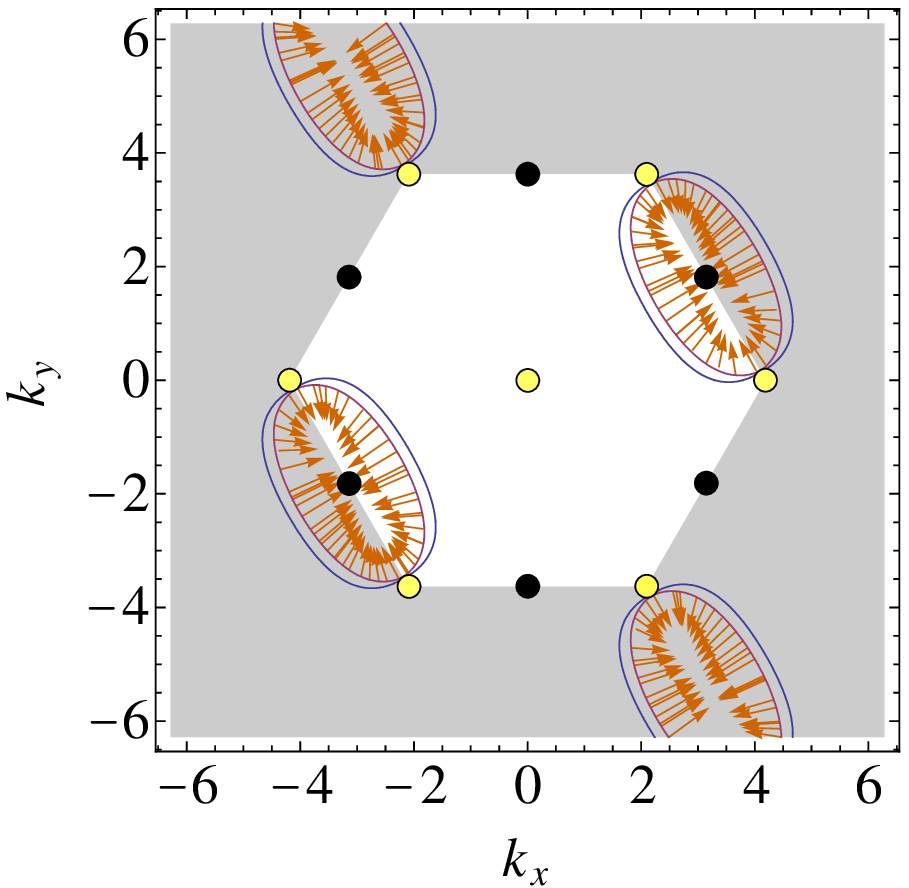}}}
   \caption{(Color online) Fermi surface for both spin-split bands (blue: $E_+$, according to Fig.\,\ref{plot:spectrum_degeneracies}) for (a) $E_F=E_{S1}-\epsilon$, (b) $E_F=E_{S1}+\epsilon$, (c) $E_F=E_{S2}-\epsilon$ and d) $E_F=E_{S2}+\epsilon$ for $\tilde t_0/t_0=0$ and $t^R=0.1t_0$. The arrows indicate ${\bf v}\sim{\bf v}^{(0)}$.}\label{plot:FermiSurf2}
    \end{center}
\end{figure}

In our case we have found that ${\bf \mathcal{A}^+(k)}={\bf \mathcal{A}^-(k)}$. Indeed, the eigenvectors, Eq.\,\ref{EV_general}, do not depend on the hopping anisotropy and therefore the Berry connection is independent of $\tilde t_0/t_0$. The Berry connection is shown as a stream plot in Fig.\,\ref{plot:Berry_connection}. Examining it in the vicinity of the degeneracy points of the energy spectrum, $\Gamma$ and $P_i$ $(i=1,\ldots,5)$, we find different winding directions for $\mathcal{A}$. It diverges exactly at these points.
In order to relate our findings to the  Berry connection we reformulate the SHC, Eq.\,(\ref{SHC}), in terms of it,\cite{2005cond.mat..4353M,PhysRevB.79.035323}
\begin{align}
 \sigma_{\kt{SH}}(E_F)={}&\frac{e}{2}\sum_{{\bf k},m=\pm}\frac{f_{E_F}(E_{m}({\bf k}))}{E_{m}({\bf k})-E_{-m}({\bf k})}\nonumber\\
{}&\cdot [{\bf \mathcal{A}}^m({\bf k})\times {\bf v^{(0)}(k)}]_z,\label{SHC_with_Berry_Connection}
\end{align}
where ${\bf v^{(0)}(k)}$ is the velocity in absence of SOC.\\ 
Let us show that the SHC sign is connected to that of $[{\bf \mathcal{A}}^m({\bf k})\times {\bf v^{(0)}(k)}]_z$.
For that purpose we first consider the isotropic case.
\begin{figure}[t]
\begin{center}
   \subfigure[\!\!\!\!\!\!\!\!\!\!\!\!\!\!\!\!\!\!]{\scalebox{0.85}{\includegraphics[width=\columnwidth*5/7]{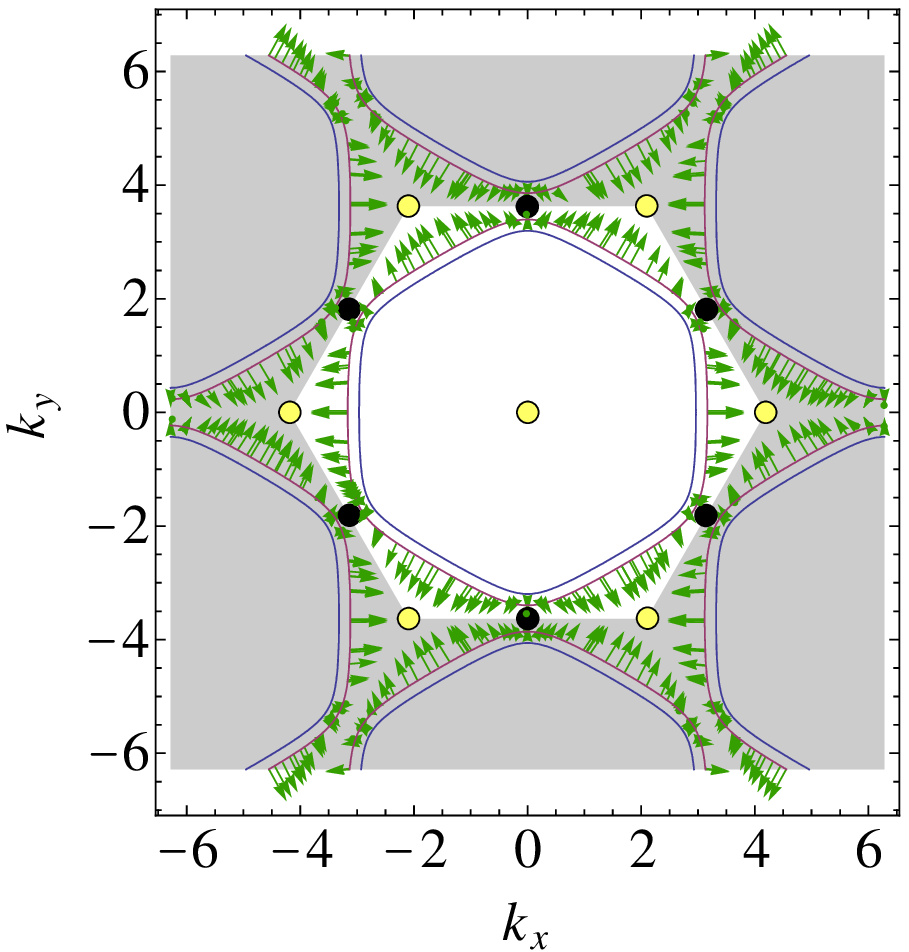}}}
   \subfigure[\!\!\!\!\!\!\!\!\!\!\!\!\!\!\!\!\!\!]{\scalebox{0.85}{\includegraphics[width=\columnwidth*5/7]{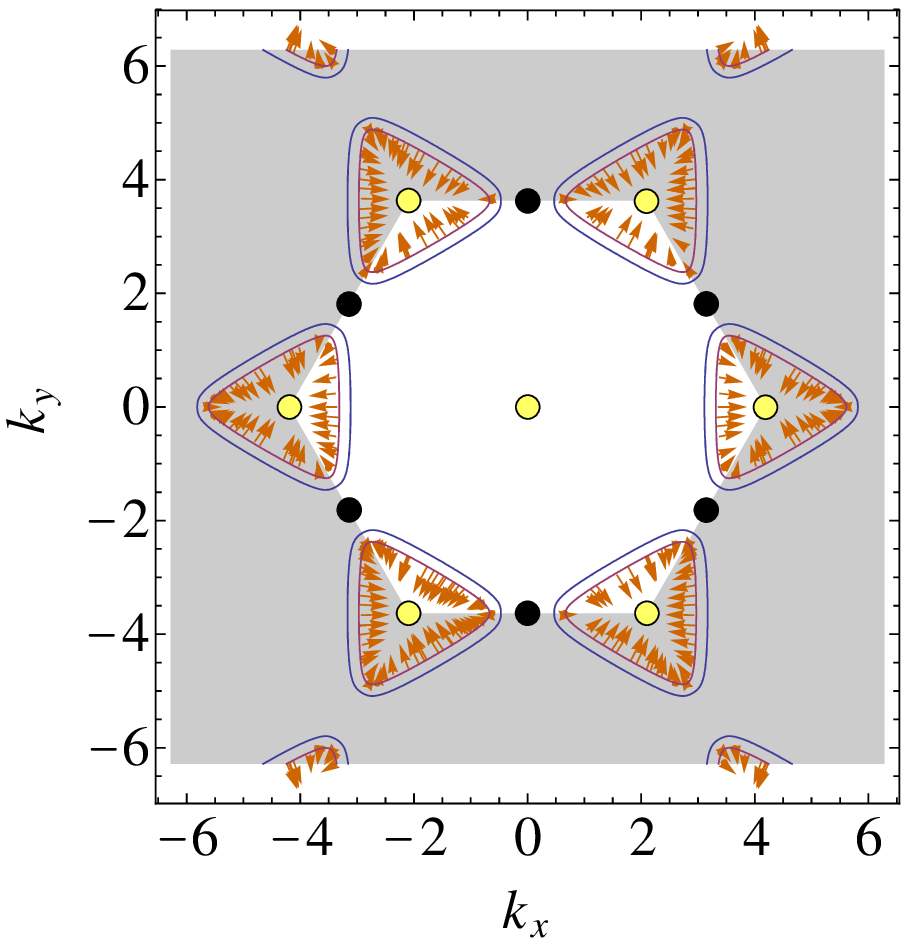}}}
   \caption{(Color online) Fermi surface for both spin-split bands (blue: $E_+$, according to Fig.\,\ref{plot:spectrum_degeneracies}) for (a) $E_F=E_{S1}-\epsilon$ and (b) $E_F=E_{S1}+\epsilon$ for $\tilde t_0=t_0$ and $t^R=0.1t_0$. The arrows indicate ${\bf v}\sim{\bf v}^{(0)}$.}\label{plot:FermiSurf1}
    \end{center}
\end{figure}
Notice that below $E_{S1}$, as the term $(E_{+}({\bf k})-E_{-}({\bf k}))$ is always positive, the sign change of SHC is associated to that of $[{\bf \mathcal{A}}^m({\bf k})\times {\bf v^{(0)}(k)}]_z$. For $E_F<E_{S1}$ the winding direction is always clockwise for the Berry connection and the velocity is always pointing outside (see Fig.\,\ref{plot:FermiSurf1} (a)), resulting then in a positive SHC.
Beyond $E_{S1}$ the Fermi surface topology is completely different, it consists now of closed loops around the $P_2$ and $P_4$ as illustrated in Fig.\,\ref{plot:FermiSurf1} (b). Furthermore, the direction of the velocity is now directed inside (towards $P_2$ and $P_4$). However, the winding directions (see Fig.\,\ref{plot:Berry_connection}) of the Berry connection is unchanged, clockwise as around the $\Gamma$ point. This results in a sign change of the spin Hall conductivity.
We now proceed further and discuss the presence of anisotropy. Since as mentioned before, $\mathcal{A}$ is independent of $\tilde t_0$ thus the changes in SHC are associated with the Fermi surface topology only.
Let us start with the extreme case of a fully anisotropic system, e.g. $\tilde t_0=0$.  As seen before, the SHC exhibits two changes of sign, one at $E_{S1}$ and the other at $E_{S2}$. Thus, we focus our attention on these two Fermi energies.
The situation at $E_{S1}$, Fig.\,\ref{plot:FermiSurf2} (a) and (b), is very similar to that of the isotropic case, except the fact that the Fermi surface at $E_F= E_{S1}+\epsilon$ is now also surrounding the point $P_5$ for which the winding direction of Berry connection is now negative (anti-clockwise). 
As the velocity direction is still pointing inwards, the negative winding around $P_5$ can lead to a sign change in $[{\bf \mathcal{A}}^m({\bf k})\times {\bf v^{(0)}(k)}]_z$, thus, to the appearance of terms in Eq.\,\ref{SHC_with_Berry_Connection} which reduce the absolute value of $\sigma_{\kt{SH}}$. This explains why $\sigma_\kt{SH}$ varies from $\sim -1.6 e/(8\pi)$ in the isotropic case to $\sim -0.5 e/(8\pi)$ for  $\tilde t_0=0$ . 
Concerning the second sign change, we find at $E_F=E_{S2}\pm\epsilon$ the same velocity direction. However, the Fermi surface at $E_F=E_{S2}+ \epsilon$ is, in contrast to Fig.\,\ref{plot:FermiSurf2} (b) and (c), surrounding only $P_5$ which as mentioned before has the opposite winding direction  with respect to that of $P_2$ and $P_4$. This yields the topological explanation of the change of sign at $E_{S2}$.
\section{Conclusion}
In this paper we have performed a detailed theoretical study of the spin Hall conductivity on the anisotropic triangular lattice in the presence of Rasbha spin-orbit coupling.
Note that experimentally anisotropy could be e.g. tuned by taking advantage of lattice mismatches in semiconductor heterostructures.
To conclude, it has been shown that on such a lattice the tight binding model should be carefully derived from the continuous Hamiltonian.
In addition, our calculations have revealed that anisotropy in the hopping term has drastic effects on the SHC especially in the hole doped regime. Furthermore, depending on the anisotropy strength, one finds either a single or two changes of sign in the SHC at particular carrier densities. The origin of the observed changes in the SHC has been interpreted and understood geometrically in terms of Berry phases. The high symmetry points which have different winding directions of the Berry connection control the rapid variation of the SHC.
More specifically, the changes of sign and amplitude of the SHC have been associated to the topology of the Fermi surface, the number of high symmetry points enclosed in it and to the sign of the velocity in the absence of SO coupling. 

\section{Acknowledgment}
This research was supported by DFG Project KE 807/6-2 \textit{Itinerant Spin Dynamics}.
\bibliographystyle{apsrev}
\bibliography{MyCollection.bib}
\end{document}